\documentclass[onecolumn,showpacs,amsmath,amssymb,aps]{revtex4-1}
\usepackage{graphicx}
\usepackage{multirow}
\usepackage{bm}
\usepackage[breaklinks=true,colorlinks=true,linkcolor=blue,urlcolor=blue,citecolor=blue]{hyperref}
\usepackage{booktabs}

\begin{document}

\title{Transient spin current under a thermal switch}
\author{Xiaobin Chen$^{1,2,*}$}
\email{Email:chenxiaobin@hit.edu.cn}
\author{Jiangtao Yuan$^{2,3}$}
\author{Gaomin Tang$^{2,3}$}
\author{Jian Wang$^{2,3,*}$}
\email{Email:jianwang@hku.hk}

\author{Zhaohui Zhang$^4$, Can-Ming Hu$^4$}
\author{Hong Guo$^{5,6}$}

\affiliation{
$^1$School of Science, Harbin Institute of Technology, Shenzhen 518055, China\\
$^2$Department of Physics and the Center of Theoretical and Computational Physics, The University of Hong Kong, Hong Kong, China \\
$^3$The University of Hong Kong Shenzhen Institute of Research and Innovation, Shenzhen 518053, China\\
$^4$Department of Physics and Astronomy, University of Manitoba, Winnipeg R3T 2N2, Canada\\
$^5$College of Physics and Energy, Shenzhen University, Shenzhen 518060, China\\
$^6$Department of Physics, 3600 University, McGill University, Montreal, Quebec H3A 2T8, Canada
}

\date{\today}

\begin{abstract}
In this work, we explore the possibility of enhancing a spin current under a thermal switch, $i.e.$, connecting the central transport region to two leads in individual thermal equilibrium abruptly. Using the nonequilibrium Green's function method for the transient spin current, we obtain a closed-form solution, which is applicable in the whole nonlinear quantum transport regime with a significant reduction of computational complexity. Furthermore, we perform a model calculation on a single-level quantum dot with Lorentzian linewidth. It shows that the transient spin current may vary spatially, causing spin accumulation or depletion in the central region.
Moreover, general enhancement of the spin current in the transient regime is observed. In particular, the in-plane components of the transient spin current may increase by 2$\sim$3 orders of magnitude compared to the steady-state thermoelectric spin current under a temperature difference of $30$~K. Our research demonstrates that ultrafast enhancement of spin currents can be effectively achieved by thermal switches.
\end{abstract}

\maketitle

\section{Introduction}

   With the ever-lasting minimization of electronic devices, reducing power consumption and dissipation of devices has become vital. This issue can be solved by using spintronic devices, which have the potential advantages of faster data processing speed, lower power consumption, and higher integration densities\cite{WolfSA_Science_2001,Kuntal_JPD2014}. For realizing spintronic devices, a major challenge is to control and manipulate spin currents, which are useful for spin injection into semiconductors\cite{ZhangShoucheng_Sci2003,Schmidt_JPD2005,Moussy_JPD2013} and magnetic reorientation of ferromagnets\cite{Slonczewski_PRB_1989,AlbertFJ_APL2000,Choi_PRL2007,cxb_PRB2017_STT}.

   Traditionally, spin currents can be manipulated using magnetic fields, bias voltages, gate voltages, $etc.$
   Besides regular electrical methods, the burgeoning research field of ``spin caloritronics'' offers a new way to manipulate spin currents, $i.e.$, thermal manipulation of spin currents\cite{Bauer_NatMater_2012,Slachter_NatPhy_2010}. Interestingly, experiments performed by Cahill $et~al.$ provided evidence that spin transfer torques, which are components of a spin current absorbed at an interface, could be enhanced by the intense and ultrafast heat current created by laser light\cite{Cahill_NatPhys2015}. Due to the simultaneous variation of the temperature, it is not clear about whether there is an intrinsic enhancement of spin currents in the transient regime or not. Although electrically-induced transient phenomena are well investigated for charge currents\cite{Joseph_PRB_2006,WangB_PRB_2010,zhanglei2012,ZhangLei_PRB2013dftTransient,ZhouChenyi_PRB2016} and spin currents\cite{Koopmans_NatComm_2014,cxb_PRB2017_STT,Zhizhou2017}, little is known about thermally-induced transient spin currents. 

   Previously, investigations of transient quantum transport phenomena mostly focused on electrically-driven ones. For a simplest two-probe transport system, one can assume that the system is in equilibrium before an arbitrary time $t=t_0$. Then, the nonequilibrium Hamiltonian is added, and the system is in nonequilibrium afterwards\cite{Jauho_PRB_1994,Joseph_PRB_2006}. However, when the temperature difference of two leads is involved, one cannot assume a sudden temperature variation of a lead at $t=t_0$ because the characteristic time scale for lattice dynamics is at least three orders larger than that of electrons\cite{Cahill_NatPhys2015}. Instead, the temperature difference can be set up before $t=t_0$ with leads disconnected with the central region. Therefore, to explore thermally-induced transient spin currents under a temperature difference, one may connect leads in different temperatures with the central part suddenly at $t=t_0$. This operation is called a ``thermal switch''\cite{WangJiansheng_PRB2010,Gaomin_NJP2017}. 


   In this work, we use the nonequilibrium Green's function (NEGF) method to investigate the transient spin current under a thermal switch. A closed-form solution is obtained and formulated in terms of steady-state nonequilibrium Green's functions, greatly simplifying the transient problem and facilitating further $ab~initio$ studies. Our formalism is applicable in the entire nonlinear quantum transport regime. Furthermore, we perform a model calculation for a quantum dot coupled to ferromagnetic leads with Lorentzian linewidths. We find that the transient spin current may vary spatially, causing spin accumulation or depletion in the central region. In addition, the transient spin current is enhanced a lot compared to the steady-state thermoelectric spin current: the in-plane components ($x,z$) of the spin current increase by several orders of magnitude; while, the out-of-plane component of the spin current increases by a few percent. Further analysis reveals the dependence of transient spin currents on the temperature, the QD energy level, and the noncollinear angle. Our studies demonstrate that spin currents can be effectively manipulated by thermal switches.

\section{Theory}
    \begin{figure}
    \centering
      \includegraphics[width=0.35\textwidth,bb=0 0 458 478]{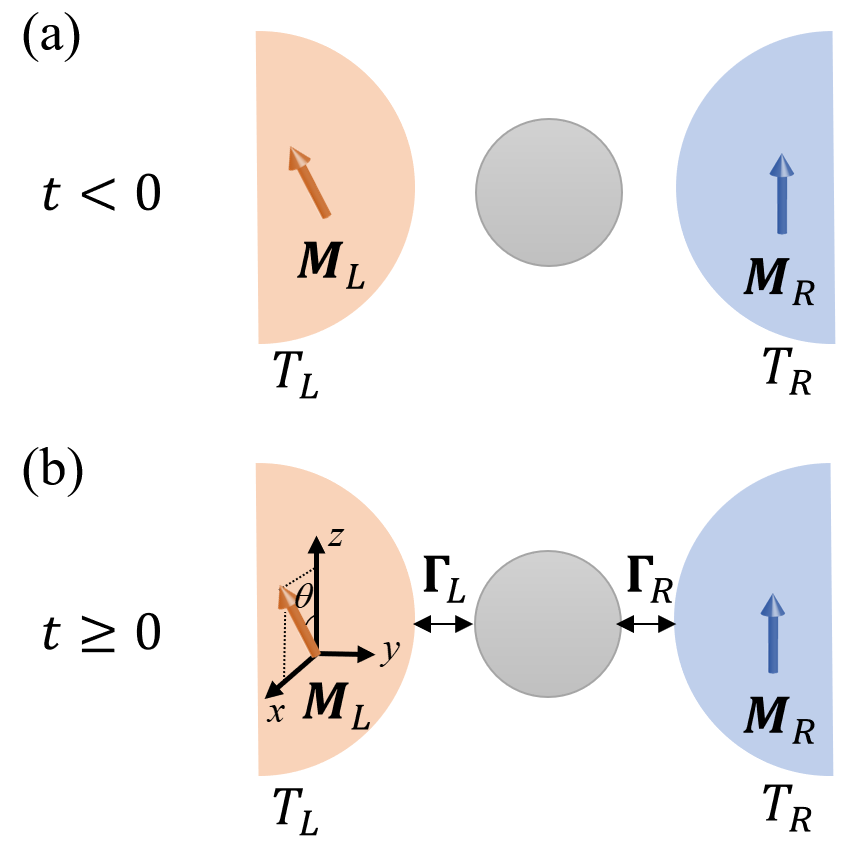}\\
      \caption{Schematic plot of the magnetic tunneling system investigated. (a) When $t<0$, leads $L$ and $R$ are in thermal equilibrium with temperatures $T_L$ and $T_R$, respectively. (b) The connection between the leads and the quantum dot is switched on at $t=0$ suddenly and persists afterwards.}\label{figModel}
    \end{figure}

   \subsection{Model Hamiltonian}
     Considering a magnetic tunnel junction (MTJ) consisting of an insulating central part ($C$) and two ferromagnetic leads ($L,R$), we can write down the Hamiltonian of the system as\cite{cxb_PRB2017_STT,cxb_prb2013_ac}
        \begin{eqnarray}
            {\hat{H}}_L = &\sum\limits_{k,s=\pm 1 }   \left\{ \left[ \varepsilon _{k L s  }+ s  {M_L}\cos \theta  \right]   \hat{c}_{k L s  }^\dag {\hat{c}}_{k L s  } \right.   \cr
            & \left. + {M_L}\sin \theta \hat{c}_{k L s  }^\dag \hat{c}_{k L \bar s }\right\},  \\
             {\hat{H}_{R}} =& \sum\limits_{ks  } {\left[ {{\varepsilon _{k {R} s  }} + s  {M_{R}}} \right] \hat{c}_{k {R} s  }^\dag { \hat{c}_{k {R} s  }}} ,\\
            {\hat{H}_{C}} =& \sum\limits_{m,s  } {{\varepsilon _m} \hat{d}_{ms  }^\dag { \hat{d}_{ms  }}}, 
            \label{eq:Hc}\\
            \hat{V} =& \sum\limits_{ s  ,m; k\alpha  \in {L,R}} {{t_{k\alpha ,m}} \hat{c}_{k\alpha s  }^\dag {\hat{d}_{ms  }} + \rm{H.c.}} ,
        \end{eqnarray}
  where $M_{L(R)}$ is the total magnetic moment of lead $L(R)$ with unit magnetization vector ${\bf{\hat M}}_L ({\bf{\hat M}}_R)$, $\varepsilon _{k L(R) s  }$ is an energy level with band index $k$ and spin $s$ in lead $L(R)$, $s=\pm 1$, and $\bar{s}=-s$.
  In addition, $\hat{c}_{k \alpha s  }^\dag (\hat{c}_{k \alpha s  }) $ creates (annihilates) an electron labeled by $k$ and $s$ in lead $\alpha$. $\hat{d}_{ms  }^\dag (\hat{d}_{ms  })$ creates (annihilates) an electron with spin $s$ and energy $\varepsilon _m$ in the central region ($C$), which is spin-degenerate. 
  Hopping between $C$ and lead $\alpha$ is described by ${t_{k\alpha ,n}}$, which is spin-independent. ``H. c." means Hermitian conjugate. In this model, ${\bf{\hat M}}_R$ aligns with the $z$-axis, ${\bf{\hat M}}_L$ lies in the $x-z$ plane, forming a noncollinear angle of $\theta$ ($0^\circ \le \theta<360^\circ$) with ${\bf{\hat M}}_R $ [see Figure 1]. In experiments, the magnetization direction of ferromagnetic leads can be manipulated by applying a magnetic field or using pinning layers.\cite{WolfSA_Science_2001,Zhaohui}

   \subsection{Transient spin currents at $t>0$ under a sudden switch-on of connection at $t=0$}
   When the connection of leads to the central region is suddenly switched on at $t=0$, the spin current flowing from the central region to lead $l$ after $t=0$ is
   \begin{align}\label{app:Jspin0t}
   J_{l;\alpha }^{spin}(t) &=  - \int_0^t {d{t_1}} {\rm{TrRe}}\left\{ {{\bf{G}}^r}\left( {t,{t_1}} \right){\bf{\Sigma }}_l^ < \left( {{t_1},t} \right){\sigma _\alpha }\right. \cr
   &\left. + {{\bf{G}}^ < }\left( {t,{t_1}} \right){\bf{\Sigma }}_l^a\left( {{t_1},t} \right){\sigma _\alpha } \right\}
   \end{align}
   according to Eq.~(\ref{app:Jspinl}). Here, $\sigma_\alpha$ is a Pauli matrix, ${{\bf{G}}^{r(<)}}$ is the retarded (lesser) Green's function of the system, and ${\bf{\Sigma }}_l^ {< (a)}$  is the lesser (advanced) self-energy of lead $l$. To obtain $J_{l;\alpha }^{spin}(t)$, one needs to obtain the Green's functions ${{\bf{G}}^ {r,<} }$ and self-energies ${\bf{\Sigma }}_l^ {a,<}$.

        Self-energies for lead $l$ are
        \begin{align}
        \Sigma _{l;n''s'',ns'}^\gamma \left( {{t_1},t} \right)
         = \theta \left( {{t_1}} \right)&
         \theta \left( t \right) \sum\limits_{k''\alpha '',k\alpha  \in l}  {} {t_{n'',k''\alpha ''}} \cdot \cr
         &g_{k''\alpha ''s'',k\alpha s'}^\gamma \left( {{t_1},t} \right){t_{k\alpha ,n}},
        \end{align}
        where $\theta(t)$ is the Heaviside step function and $\gamma=r,a,>,<$.
        When both time variables are larger than 0, we can also obtain the Fourier transform of the self-energies for the connected system as\cite{footnote1}$({t , t_1 > 0})$:
        \begin{align}\label{app:sigmar_Fourier}
        \Sigma _{l;n''s'',ns'}^\gamma \left( {{t_1},t} \right) = \int {\frac{{d\varepsilon }}{{2\pi }}} \Sigma _{l;n''s'',ns'}^\gamma \left( \varepsilon  \right) {e^{ - {\rm{i}}\varepsilon \left( {{t_1} - t} \right)}} ,
        \end{align}
        where
        \begin{align}\label{eq:selfl}
        \Sigma _{l;n''s'',ns'}^\gamma \left( \varepsilon  \right) = \sum\limits_{k\alpha ,k''\alpha '' \in l}  {{t_{n'',k''\alpha ''}}g_{k''\alpha ''s'',k\alpha s'}^\gamma \left( \varepsilon  \right){t_{k\alpha ,n}}}.
        \end{align}
        Thus, self-energies actually depend on time difference when both time variables $t,t'$ are later than 0 $\left( {t, t_1 > 0} \right)$, $i.e.$,
            \begin{align}
            {\bf{\Sigma }}_l^\gamma \left( {{t_1},t} \right) = {\bf{\Sigma }}_l^\gamma \left( {{t_1} - t} \right).
            \end{align}
        After obtaining the steady-state self-energies in Eq.~(\ref{eq:selfl}), the double-time self-energies can be determined using Eq.~(\ref{app:sigmar_Fourier}).

        For the lesser and retarded Green's functions of the central system, we note the Keldysh formula\cite{Jauho_Book,jiangtao,TangGaomin_prb2014}
        \begin{align}
        {{\bf{G}}^ < } = {\rm{(1 + }}{{\bf{G}}^r}{{\bf{\Sigma }}^r}){\bf{G}}_0^ < \left( {1 + {{\bf{\Sigma }}^a}{{\bf{G}}^a}} \right) + {{\bf{G}}^r}{{\bf{\Sigma }}^ < }{{\bf{G}}^a},
        \end{align}
        where a product of two terms is interpreted as a matrix product in the internal variable time. With the first term on the right-hand side rewritten, this equation can be further simplified to be\cite{Stefanucci_PRB2004}
        \begin{align}
        {{\bf{G}}^ < }\left( {t,t'} \right) &={{\bf{G}}^r}\left( {t,0} \right){\bf{G}}^ < \left( {0,0} \right){{\bf{G}}^a}\left( {0,t'} \right)+\cr
        & \iint  _{ - \infty }^{ + \infty } {d{t_1}}{d{t_2}}
        {{\bf{G}}^r}\left( {t,{t_1}} \right){{\bf{\Sigma }}^ < }\left( {{t_1},{t_2}} \right){{\bf{G}}^a}\left( {{t_2},\;t'} \right), \cr
        \end{align}
        where ${\bf{G}}^ < \left( {0,0} \right)$ is the initial population of the central region at $t=0$. From this equation, ${{\bf{G}}^ < }$ can be obtained once ${{\bf{G}}^ r }$ and the initial population are known. For using a thermal switch at $t=0$, we have\cite{jiangtao}
        \begin{align}\label{app:keldysh}
        {{\bf{G}}^ < }\left( {t,t'} \right) &={{\bf{G}}^r}\left( {t,0} \right){\bf{G}}^ < \left( {0,0} \right){{\bf{G}}^a}\left( {0,t'} \right)+\cr
        & \iint  _{ 0 }^{ + \infty } {d{t_1}}{d{t_2}}
        {{\bf{G}}^r}\left( {t,{t_1}} \right){{\bf{\Sigma }}^ < }\left( {{t_1},{t_2}} \right){{\bf{G}}^a}\left( {{t_2},\;t'} \right). \cr
        \end{align}
        Then, for the retarded Green's function of the central region, we may utilize the Dyson equation
        \begin{align}\label{app:dyson}
        {{\bf{G}}^r}\left( {t,t'} \right) = {\bf{G}}_0^r & \left( {t,t'} \right) + \int_0^{ + \infty }  \int_0^{ + \infty } {d{t_1}}{d{t_2}}  \cr
        &{\bf{G}}_0^r\left( {t,{t_1}} \right){{\bf{\Sigma }}^r}\left( {{t_1},{t_2}} \right){{\bf{G}}^r}\left( {{t_2},\;t'} \right),
        \end{align}
        where ${\bf G}_0^r(t,t')$ is the retarded Green's function of the disconnected central region and it depends on $t-t'$ :
        \begin{align}\label{app:G0r(E)}
        {\bf{G}}_0^r\left( {t,t'} \right) = {\bf{G}}_0^r\left( {t - t'} \right) = \int   {\frac{{d\varepsilon} }{{2\pi }}{\bf{G}}_0^r\left( \varepsilon \right)} {e^{ - \textrm{i}\varepsilon\left( {t - t'} \right)}}.
        \end{align}
        By introducing a double-time Fourier transformation of Green's functions, we can prove that [see \ref{app:proof}]:
         \begin{align}\label{eq:Gr=GbarR}
        {{\bf{G}}^r}\left( {t,t'} \right) = {\bf{\bar G}}^r\left( {t-t'} \right), \quad\quad t,t'>0,
        \end{align}
        where $ {\bf{\bar G}}^r\left( t,t' \right)$ is the retarded Green's function in the steady-state limit. Its Fourier component ${\bf{\bar G}}^r\left( \varepsilon \right)$ can be calculated using
         \begin{align}\label{eq:Grbar}
        {\bf{\bar G}}^r\left( \varepsilon \right) &= {\left[ {\varepsilon + \textrm{i}\eta  - {{\bf{H}}_0} - {\bf{\Sigma }}}^r \right]^{ - 1}},
        \end{align}
        where $\eta$ is an infinitesimal positive number. Although the system undergoes a sudden change at $t=0$, the retarded Green's function ${{\bf{G}}^r}\left( {t,t'} \right)$, strikingly, has time-translational invariance when $t,t'>0$. This result is key to our formulas, because it means that the double-time Green's function ${\bf{G}}^r(t,t')$ can be obtained through Fourier transform of ${\bf{\bar G}}^r\left( \varepsilon \right)$. Then, the double-time lesser Green's function can be acquired using Eq.~(\ref{app:keldysh}), which is in a closed form and does not need iterative calculation.

        Finally, the $\alpha$-component of the transient spin current flowing into lead $l$ [Eq.~(\ref{app:Jspin0t})] can be computed as
       \begin{align}\label{eq:JLa_spin}
       J_{l;\alpha }^{spin}
       &=- \int_{ - \infty }^{ + \infty } {\frac{{d\varepsilon }}{{2\pi }}} {\rm{ReTr}}\left[ {\bf{A}}\left( {\varepsilon ,t} \right){\bf{\Sigma }}_l^ < \left( \varepsilon  \right){\sigma _\alpha } \right.\cr
       &+ {{\bf{A}}\left( {\varepsilon ,t} \right){{\bf{\Sigma }}^ < }\left( \varepsilon  \right){{\bf{B}}_l}\left( {\varepsilon ,t} \right){\sigma _\alpha }}  \cr
       &+ \left. {{ {\bf G}^r}\left( {t,0} \right){{\bf G}^ < }\left( {0,0} \right){{\bar {\bf G}}^a}\left( \varepsilon  \right){\bf C}_l\left( {\varepsilon ,t} \right){\sigma _\alpha }} \right],
       \end{align}
       where 
       \begin{align}
        {\bf{A}}\left( {\varepsilon ,t} \right)
        & = {\bf{\bar G}}_{}^r\left( \varepsilon  \right) + \int_{ - \infty }^{ + \infty } {\frac{{d\omega }}{{2\pi \textrm{i}}}\frac{{{e^{ - \textrm{i}\left( {\omega  - \varepsilon } \right)t}}}}{{\varepsilon  - \omega  + \textrm{i}{0^ + }}}{\bf{\bar G}}_{}^r\left( \omega  \right)}, \label{eq:A(E,t)} \\
       {{\bf{B}}_l}\left( {\varepsilon ,t} \right)
        &= {{{\bf{\bar G}}}^a}{\bf{\Sigma }}_l^a - \int_{}^{} {\frac{{d\omega }}{{2\pi \textrm{i}}}} \frac{{{e^{\textrm{i}\left( {\omega  - \varepsilon } \right)t}}}}{{\varepsilon  - \omega  - \textrm{i}{0^ + }}}{{{\bf{\bar G}}}^a}\left( \omega  \right){\bf{\Sigma }}_l^a(\omega), \label{eq:B(E,t)}\\
        {\bf C}_l\left( {\varepsilon ,t} \right) &= \int_{}^{} {\frac{{d\omega }}{{2\pi \textrm{i}}}} {\bf \Sigma} _l^a\left( \omega  \right)\frac{{ - {e^{\textrm{i}\omega t}}}}{{\varepsilon  - \omega  + \textrm{i}{0^ + }}}, \label{eq:C(E,t)}
       \end{align}
       respectively. In particular, when the central part is initially unpopulated, the third term in Eq.~(\ref{eq:JLa_spin}) is zero. This closed-form formula for the transient spin current greatly simplifies numerical calculations, which would be rather complicated and time-consuming using the perturbation theory\cite{WangJiansheng_PRB2010}. 
       So far, we have solved out the transient spin current and the final expression [Eq.~(\ref{eq:JLa_spin})] requires only the steady-state quantities. We shall apply Eqs.~(\ref{eq:JLa_spin}-\ref{eq:C(E,t)}) to the simplest spin-degenerate single-level quantum dot model to investigate the transient spin current under a thermal switch.


\section{Spin-degenerate single-level QD with Lorentzian bandwidth functions}
       In the above section, the transient spin current under a thermal switch is expressed as one single integral over energy. In some cases where self-energies of leads are in simple analytic forms, the retarded Green's function of the central region and even the ${\bf{A}}(\varepsilon,t)$ and ${\bf{B}}_l(\varepsilon,t)$ functions can be acquired analytically.
    \begin{figure*}
      \centering
      \includegraphics[width=1\textwidth,bb=0 0 864 216]{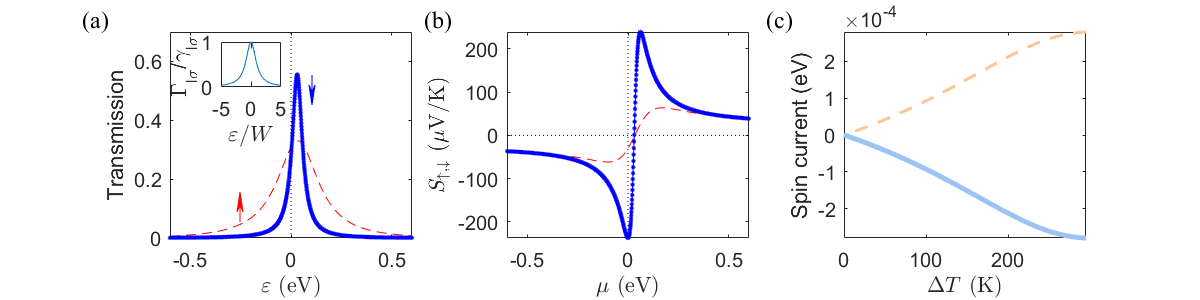}\\
      \caption{Transport properties of the FM/QD/FM system in the long-time steady-state limit. (a) Transmission and (b) Seebeck coefficients of spin-up (red dash line) and spin-down (thick blue solid line) channels as functions of energy $\varepsilon$ and chemical potential $\mu$, respectively, when $T_L=T_R=300$ K. Lorentzian line-shape bandwidth function as a function of $\varepsilon/W$ is also shown in the inset of (a). (c) Spin currents in lead $L$ (thin light-orange dashed line) and $R$ (thick light-blue line) as functions of temperature difference ${\Delta}T=T_L-T_R$, where $T_L$ is fixed to be 300 K. Here, $\gamma_{L\uparrow},\gamma_{L\downarrow},\gamma_{R\uparrow},\gamma_{R\downarrow}$ are chosen to be 0.2, 0.05, 0.02, and 0.01~eV, respectively. Other parameters are set as: $W=1$~eV, $\varepsilon_0 =0.03$~eV, $\eta=1\times 10^{-13}$ eV, $\theta=0^\circ$.}\label{figTrans}
    \end{figure*}

       We consider a system which consists of two ferromagnetic (FM) leads and a single-level quantum dot (QD) between the leads. Similar FM/QD/FM systems have been widely studied previously\cite{Datta_2005_transistor,Swirkowicz_PRB_2009,cxb_prb2013_ac,Tang_Arxiv2017}. When $t<0$, three parts are disconnected and in individual thermal equilibrium. When $t=0$, the QD is connected to two leads suddenly. [See Fig.~\ref{figModel}] Coulomb blockade effects are usually found in single-level quantum dots where the single-electron charging energy $U_0$ exceeds the thermal energy $k_BT$ and the level broadening ($\gamma_{\uparrow,\downarrow}$ in below). Therefore, we assume that the wavefunction in the single-level QD model is well-delocalized, having $U_0$ small enough, and thus the Coulomb blockade effects can be neglected. Also, defined as
       \begin{align}
       {\Gamma _l}{\rm{ =  - 2Im}}{\Sigma _l^r },
       \end{align}
        the bandwidth function of lead $l$ (${\Gamma _l}$) is a key factor for transport properties. To obtain analytical results, the bandwidth function is usually supposed to be constant (wide-band limit).\cite{Jauho_PRB_1994,cxb_prb2013_ac,ZhouChenyi_PRB2016} In our work, we go beyond the wide-band limit and introduce the Lorentzian linewidths. We suppose
       that the bandwidth function of lead $l$ in parallel configuration ($\theta=0$) is in Lorentzian line-shape as\cite{Joseph_PRB_2006,jiangtao} [the inset of Fig.~\ref{figTrans}(a)]
       \begin{align}
       {\Gamma _{l\sigma }}(\varepsilon) = \frac{{{\gamma _{l\sigma }}{W^2}}}{{{\varepsilon ^2} + {W^2}}},
       \end{align}
       where $\sigma$($=\uparrow,\downarrow$) labels spin ($\left|z\uparrow\right>$, $\left|z\downarrow \right>$) states, ${\gamma _{l\sigma }}$ is the linewidth amplitude of spin-$\sigma$ channels in lead $l$, and $W$ is the bandwidth. Lorentzian linewidths are mathematically convenient and are widely used for introducing finite-bandwidth effects. Then, the retarded self-energies of leads $L$ and $R$ can be obtained by utilizing the spectral representation of Green's functions as\cite{Joseph_PRB_2006}
        \begin{align}
        \Sigma _{l\sigma }^{r,a} (\varepsilon)& = \int_{}^{} {\frac{{d\omega }}{{2\pi }}\frac{{{\Gamma _{l\sigma }}\left( \varepsilon  \right)}}{{\varepsilon  - \omega  \pm \textrm{i}{0^ + }}} = } \frac{1}{2}\frac{{{\gamma _{l\sigma }}W}}{{\varepsilon  \pm \textrm{i}W}}, \\
        {\bf{\Sigma }}_{L0}^r\left( \varepsilon  \right)
        & = \frac{1}{2}\frac{W}{{\varepsilon  + \textrm{i}W}}\left( {\begin{array}{*{20}{c}}
        {{\gamma _{L \uparrow }}}&{}\\
        {}&{{\gamma _{L \downarrow }}}
        \end{array}} \right), \\
        {\bf{\Sigma }}_R^r\left( \varepsilon  \right)
        &= \frac{1}{2}\frac{W}{{\varepsilon  + \textrm{i}W}}\left( {\begin{array}{*{20}{c}}
        {{\gamma _{R \uparrow }}}&{}\\
        {}&{{\gamma _{R \downarrow }}}
        \end{array}} \right).
        \end{align}
        For a general noncollinear angle $\theta$, the retarded self-energy of the left lead can be written as
        \begin{align}
         {\bf{\Sigma }}_{L}^r = {R^ \dagger } {\bf{\Sigma }}_{L0}^r R
        \end{align}
        with the rotation matrix\cite{cxb_prb2013_ac}
        \begin{equation}
        R = \left( {\begin{array}{*{20}{c}}
        {\cos \frac{\theta }{2}}&{\sin \frac{\theta }{2}}\\
        { - \sin \frac{\theta }{2}}&{\cos \frac{\theta }{2}}
        \end{array}} \right).
        \end{equation}
       With well-defined self-energies, the transient spin current can then be obtained. Analytical results of the auxiliary functions ${\bf{A}} \left( {\varepsilon ,t} \right)$ and ${{\bf{B}}_l}\left( {\varepsilon ,t} \right)$ for this single-level QD model can be found in \ref{app:AB}.

    Before investigating transient behaviors, we would like to present transport properties of the steady-state limit at $t\to \infty$, where the spin current is driven by the temperature gradient. Steady-state limit restrains the long-time tail, which is necessary for us to get a full picture of the transient behavior. Due to the retarded nature of Green's functions, all time-related quantities in ${\bf{A}}(\varepsilon,t)$ and ${\bf{B}}_l(\varepsilon,t)$ vanish when $t\to \infty$, which means that only the dc component persists to infinite time. Assuming that the initial population on the central QD is zero, we can prove that\cite{cxb_PRB2017_STT} 
    \begin{equation}
    J_{L;x/z}^{dc;spin} =  - \frac{1}{2}\int_{ - \infty }^{ + \infty } {\frac{{d\varepsilon }}{{2\pi }}} \left( {{f_L} - {f_R}} \right){\rm{Tr}}\left[ {{{{\bf{\bar G}}}^r}{{\bf{\Gamma }}_R}{{{\bf{\bar G}}}^a}{{\bf \Gamma} _L}{\sigma _{x/z} }} \right]
    \end{equation}
    and a similar form for $J_{R;x/z}^{dc;spin}$. For collinear spin systems with spin polarization along $z$, it is further reduced to
    \begin{equation}
    J_{L;z}^{dc;spin} =  - \frac{1}{2}\int_{ - \infty }^{ + \infty } {\frac{{d\varepsilon }}{{2\pi }}} \left( {{f_L} - {f_R}} \right)\left[ {{\mathcal{T}_ \uparrow }\left( \varepsilon  \right) - {\mathcal{T}_ \downarrow }\left( \varepsilon  \right)} \right].
    \end{equation}
    Actually, the dc spin current is proportional to the dc charge current\cite{Jauho_PRB2009}. This connection can be attributed to the fact that the investigated spin angular momentum is carried by electrons. Under a thermal voltage, the dc spin current is the thermoelectric spin current, which within linear response theory can be expressed as\cite{cxb_PRB2015valley}
    \begin{equation}\label{eq:GST}
    J_{L;z}^{dc;spin} =\frac{\hbar }{{2e}} \left( {{G_ \uparrow }{S_ \uparrow } - {G_ \downarrow }{S_ \downarrow }} \right)\Delta T,
    \end{equation}
    where $\Delta T=T_L-T_R$, and $G_\sigma$ ($S_\sigma$) is the electrical conductance (Seebeck coefficient) of spin-$\sigma$ electrons.

\section{Results and Discussion}
    To begin with, we choose a particular set of parameters to investigate the transient spin current in detail and focus on the parallel configuration, $i.e.$, $\theta=0$, where only spin-$z$ current exists. Considering that the bandwidth function in a Phenyldithiol molecular junction could be $\Gamma=0.11$ eV or $0.0042$~eV\cite{Datta_prb_2003}, we choose
     $\gamma_{L\uparrow}=0.2$, $\gamma_{L\downarrow}=0.05$, $\gamma_{R\uparrow}=0.02$, and $\gamma_{R\downarrow}=0.01$~eV, corresponding to lead $L$ and $R$ with spin polarization 60\% and 33\%, respectively.  The energy level of QD is $\varepsilon_0=0.03$ eV for a better thermoelectric performance, as what we shall see below. Also, we assume that the energy level of the QD is initially unoccupied.
    \begin{figure} 
    \centering
      \includegraphics[width=0.4\textwidth,bb= 0 0 350 486]{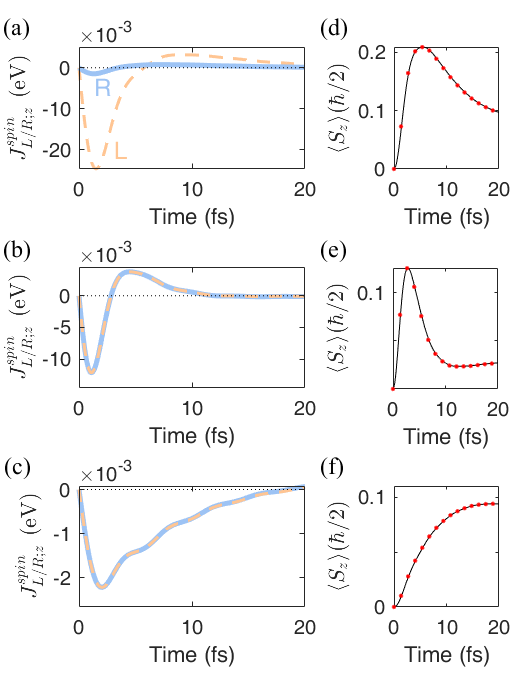}\\
      \caption{(a-c) Transient spin currents at lead lead $L$ (thin light-orange dashed line) and $R$ (thick light-blue line) as a function of time at the parallel configuration ($\theta=0^\circ$) during $t=0$ to $20$ fs. (a) Leads $L$ and $R$ are different: $\gamma_{L\uparrow}=0.2, \gamma_{L\downarrow}=0.05, \gamma_{R\uparrow}=0.02, \gamma_{R\downarrow}=0.01$ eV. (b)(c) Leads $L$ and $R$ are identical, where $\gamma_{L(R)\uparrow}, \gamma_{L(R)\downarrow}$ are set to be (b) $0.2,0.1$ eV, and (c) $0.02,0.01$ eV, respectively. (d-f) Accumulated spin angular momentum at the central region estimated as a function of time, corresponding to (a-c), respectively. Blue solid lines indicate results obtained using the accumulated number of electrons and the red dots indicate results of conservation of spin angular momentum. The other parameters are set as: $W=1$~eV, $\varepsilon_0 =0.03$~eV, $\eta=1\times 10^{-13}$ eV, $T_L=300$~K, and $T_R=270$~K.}\label{figJ}
    \end{figure}

    In Fig.~\ref{figTrans} (a), spin-resolved transmission coefficients are plotted as a function of energy. It shows that transmission spectra of spin-up and down electrons exhibit resonant tunneling peaks near $\varepsilon_0$. Both transmission peaks are in the typical Lorentzian line shape. In particular, the transmission of spin-up electrons has a wider peak than that of spin-down electrons. This phenomenon is due to the stronger coupling between the leads and the central region of the spin-up electrons in our model.\cite{Jauho_Book,Dattabook,Bulter_PRB2001}
    Heights of resonant peaks are less than 1, which can be attributed to the scattering caused by mismatched leads. The corresponding Seebeck coefficients are further shown in Fig.~\ref{figTrans}(b). The Seebeck coefficient of spin-down electrons has a larger absolute value than that of spin-up electrons within the range of $[-0.58, 0.58]$ eV, thanks to the narrow width of spin-down resonant peak\cite{cxb_prb2013_ac}. If $\varepsilon_0=0$, both Seebeck coefficients of spin-up and spin-down electrons would be zero due to the electron-hole symmetry in our model. Thus, we set $\varepsilon_0$ slightly away from the zero point and near the maximum absolute value, to have a sizable thermoelectric spin current.
    It is also shown that when $\mu=0$, both $S_\uparrow$ and $S_\downarrow$ are negative, indicative of electron-like transport. Since $S_\downarrow=-237$~$\mu$V/K and $S_\uparrow=-30.1$~$\mu$V/K at $\mu=0$, the spin current is dominated by spin-down electrons at $\mu=0$ and $S_s=S_\uparrow-S_\downarrow$ basically follows the sign of $-S_\downarrow$.

    In Fig.~\ref{figTrans} (c), dc spin current $J_{L/R}^{spin}$ is shown as a function of temperature difference. It is clearly demonstrated that $J_{L}^{spin}=-J_{R}^{spin}$, indicative of the steady-state condition and conservation of spin angular momentum. Noting that $J_{L}^{spin}>0$, $T_L>T_R$, and the system is in electron-like regime, one knows that spin current $J_{L}^{spin}$ is dominated by hot spin-down electrons transporting from $L$ to $R$. When no temperature difference is present, the spin current should be zero since no driving forces are involved. Then, as the temperature difference $\Delta T=T_L-T_R$ increases within the linear-response regime, the steady-state thermoelectric spin current increases linearly, which is in good agreement with Eq.~(\ref{eq:GST}). It is worth noting that the thermoelectric spin current is about $2.41\times10^{-5}$ eV when $\Delta T=30$~K and $T_R=270$ K.

 \begin{figure*}
      \includegraphics[width=1.0\textwidth,bb=0 0 972 410]{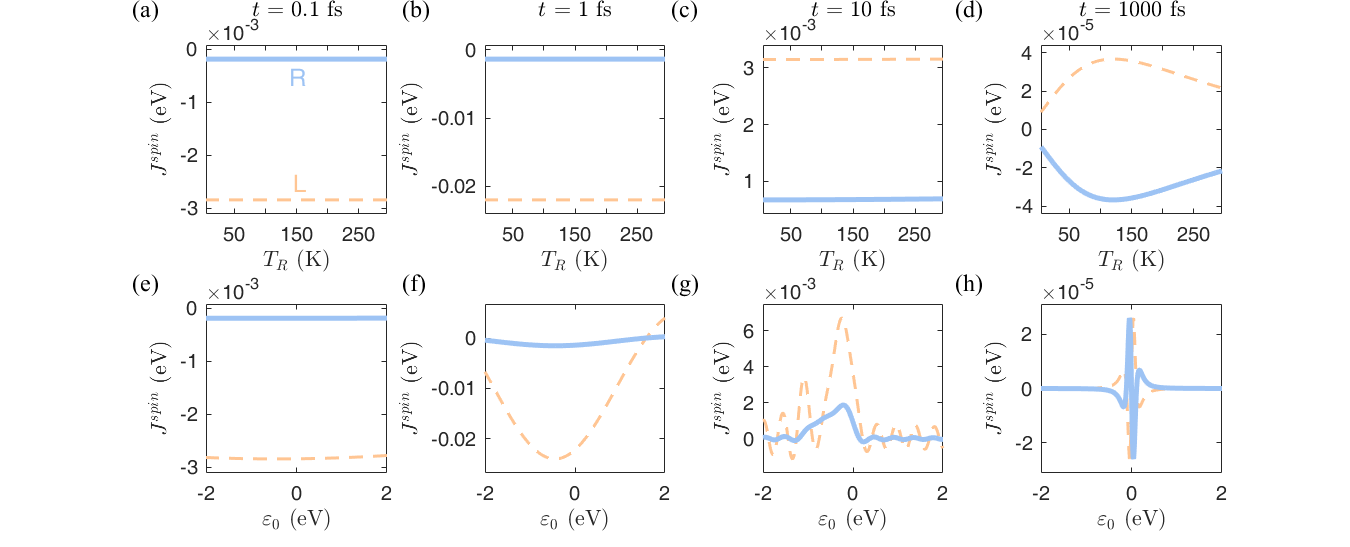}\\
      \caption{(a)-(d) Temperature and (e)-(h) $\varepsilon_0$ dependence of the transient spin current at lead $L$ (thin black line) and $R$ (thick red line) at different times: (a,e) $t=0.1$~fs; (b,f) $t=1$~fs; (c,g) $t=10$~fs; (d,h) $t=1000$~fs. In (a)-(d), $T_L=T_R+30$~K; while in (e)-(h), $T_L=300$~K, $T_R=270$~K. The other constant parameters are set to be the same with those in Fig.~\ref{figJ}(a). }\label{figTe0}
    \end{figure*}

    Now we move one step forward to obtain the transient spin current as a function of time. Results with three different parameter sets are demonstrated in Fig.~\ref{figJ}. Various coupling strength parameters are chosen to compute the transient spin current in two leads. As illustrated in Fig.~\ref{figJ}(a), the transient spin current $J_{L(R)}^{spin}$ starts from zero at $t=0$, showing reasonable consistency in time domain. Also, the transient spin current flows out of both leads immediately after $t=0$. In particular, lead $L$, which has higher spin polarization, has a larger transient spin current.
    Remarkably, the transient spin current of lead $L$ reaches $-0.0248$ eV, which is \textbf{three orders of magnitude larger} than that of the steady-state dc spin current shown in the long-time limit of spin currents when $\Delta T=30$~K [Fig.~\ref{figTrans} (c)]. The enhancement factor,
    \begin{align}\label{eq:PX}
    P(X)=\frac{|X|_{max}}{|X(t\to \infty)|},
    \end{align}
    is $P(J_{R;z}^{spin})=1.0\times10^3$.  Compared with normal enhancement of $1\sim6$ times in electrically-induced transient charge or spin currents\cite{WangB_PRB_2010,Zhizhou2017}, this result demonstrates that the transient spin current can be anomalously larger than the corresponding steady-state thermoelectric spin current.

    When two leads have the same coupling strength with the central QD, the transient spin current has nearly the same magnitude in two leads, as shown in Fig.~\ref{figJ} (b) and (c). This overlap of the transient spin current in two leads also implies negligible influence of the thermoelectric spin current near $t=0$. Also, when the coupling strength is weakened [Fig.~\ref{figJ} (c) compared to Fig.~\ref{figJ} (b)], the enhancement factor $P$ significantly decreases from $4.9\times10^2$ to $1.2\times10^2$.

    As reflected in Fig.~\ref{figTrans}, $J_L^{spin}+J_R^{spin}=0$ should be satisfied for steady-state transport. For the transient spin current, $J_L^{spin}+J_R^{spin}\ne 0$ as evidenced in Fig.~\ref{figJ}(a-c). Conservation of spin angular momentum implies that there is accumulation or depletion of spins in the central QD:
    \begin{equation}
    \int_0^t {\left( {J_{L;z}^{spin} + J_{R;z}^{spin}} \right)dt}  + \left\langle {{S_z}\left( t \right)} \right\rangle  = 0,
    \end{equation}
    where $\left\langle {{S_z}\left( t \right)} \right\rangle$ is the expected value of spin angular momentum in the QD at time $t$.
    Therefore, we may estimate the accumulation of spins using
    \begin{equation}
    \left\langle {{S_z}\left( t \right)} \right\rangle  =  - \int_0^t {\left( {J_{L;z}^{spin} + J_{R;z}^{spin}} \right)dt} .
    \end{equation}
    Alternatively, the accumulation of spins can be calculated using the number of accumulated electrons:
    \begin{equation}
    \left\langle {{S_z}} \right\rangle  = \frac{\hbar }{2}\left( {\left\langle {{n_ \uparrow }} \right\rangle  - \left\langle {{n_ \downarrow }} \right\rangle } \right),
    \end{equation}
    where $\left\langle {{n_\sigma }} \right\rangle $ is the number of spin-$\sigma$ electrons in the QD:
    \begin{equation}
    \left\langle {{n_\sigma }} \right\rangle  =  - iG_{\sigma \sigma }^ < \left( {t,t} \right).
    \end{equation}
    As demonstrated in Fig.~\ref{figJ}(d-f), two methods lead to the same results. From Fig.~\ref{figJ}(d-f), one can see that after $t=0$ spin angular momentum accumulates quickly to a maximum and then decreases. The accumulation of spins in the central region is in accordance with the time variation of the transient spin current.

    As demonstrated above, transient spin currents can be much larger than the steady-state thermoelectric spin currents. To examine the influence of temperature, transient spin currents at different times $t$ as a function of $T_R$ are plotted in Fig.~\ref{figTe0}(a)-(d). For $t=0.1, 1$ and 10~fs, the transient spin current in both leads show nearly no dependence on $T_R$. And when $t=1000$~fs, it displays the steady-state signature, where $J_L^{spin}=-J_R^{spin}$. From Fig.~\ref{figTe0}(d), variation of the spin current is up to $3\times10^{-5}$ eV. Therefore, the temperature mainly changes the dc thermoelectric spin current and has little effect on the fast transient region near $t=0$. Interestingly, the independence of the transient spin current near $t=0$ and the quasi-linear dependence of dc thermoelectric spin currents near $t\gg0$ on $\Delta T$ lead to a rough estimation that the enhancement factor varies with $\Delta T$ in a fashion $\propto 1/\Delta T$. Thus, the enhancement of $J_{R;z}^{spin}$ when $\Delta T=3$~K would be about ten times larger than that of $J_{R;z}^{spin}$ when $\Delta T=30$~K.

    By contrast, the transient spin current significantly depends on the quantum dot energy level $\varepsilon_0$, as plotted in Fig.~\ref{figTe0}(e)-(h). Although the transient spin current has only a little dependence on $\varepsilon_0$ when $t=0.1$~fs, it changes a lot when $t=1$~fs. Strong dependence is also shown at $t=10$~fs and $t=1000$~fs. For $t=10$~fs, variation is contributed by the transient component. While, for $t=1000$~fs, variation is caused by the steady-state thermoelectric component.\cite{cxb_PRB2015valley}. 

\begin{figure}
  \centering
  \includegraphics[width=0.42\textwidth,bb=0 0 402 528]{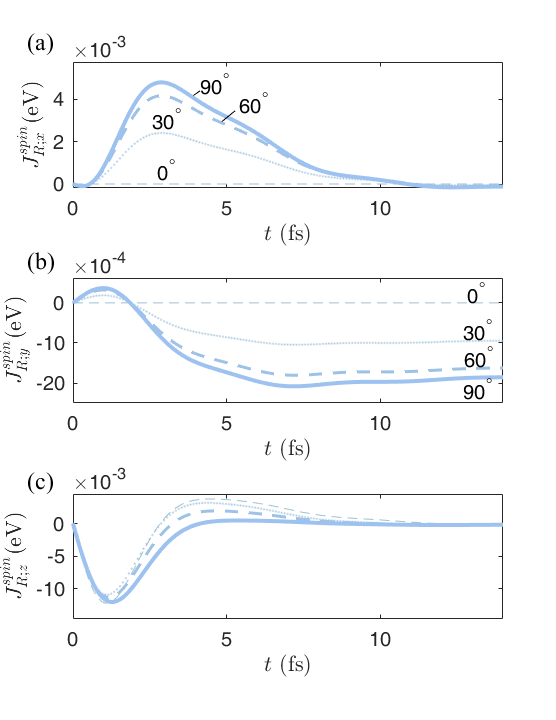}\\
  \caption{Transient behaviors of the (a) $x$, (b) $y$, and (c) $z$ components of the spin current flowing through lead $R$ at different noncollinear angles: $\theta=0^\circ$(thin dashed line), 30$^\circ$(dotted line), 60$^\circ$(thick dashed line), and 90$^\circ$ (solid line). Other parameters are set to be the same with those in Fig.~\ref{figJ}(b). }\label{fig:theta}
\end{figure}

    Finally, we shall have a short discussion about the dependence of transient spin current on the noncollinear angle $\theta$. In our model, the magnetization direction of lead $R$ is along $z$, thus the $x$($y$) component of the spin current actually corresponds to the in-plane (out-of-plane) spin transfer torque (STT), respectively\cite{cxb_PRB2017_STT}. For simplicity, we choose identical leads, $i.e.$, the same coupling parameters of leads $L$ and $R$, $\gamma_{L\sigma}$=$\gamma_{R\sigma}$, and focus on the transient spin current in lead $R$. Results are demonstrated in Fig.~\ref{fig:theta}. As mentioned above, when $\theta=0$, both $x$ and $y$ components of the spin current are zero (Fig.~\ref{fig:theta}). As $\theta$ increases from 0$^\circ$ to 90$^\circ$, the absolute values of both $J_{R;x}^{spin}$ and $J_{R;y}^{spin}$ increase from 0 to the maximum numbers.
    In addition, the angular dependence of transient $J_{R;x}^{spin}$ and $J_{R;y}^{spin}$ follow roughly the  $\sin\theta$ function, similarly to the transient STTs realized using the traditional electrical approaches\cite{Zhizhou2017}.
    For the in-plane $J_{R;x}^{spin}$, its maximum value in the transient region far exceeds the steady-state limit, which is dominated by the thermoelectric spin current. By contrast, the out-of-plane $J_{R;y}^{spin}$ has a rather large steady-state zero-field STT, and the transient enhancement is very small.
     The enhancement factor [Eq.~(\ref{eq:PX})] for the $x$ and $y$ components of the spin current are biggest at $\theta=\pi/2$, for which we have
     \begin{equation}
     P(J_{L;x}^{spin})=3.9\times10^2
     \end{equation}
     and
     \begin{equation}
     P(J_{L;y}^{spin})=1.16,
     \end{equation}
     respectively. As for the spin-$z$ current, it is well-known that the spin-$z$ current has the biggest and smallest value at $\theta=0$ and $\theta=180^\circ$, respectively. As a result, the noncollinearity angle has a much smaller impact on $J_{R;z}^{spin}$ than on other components.


   In summary, we investigated the transient spin current under a thermal switch using the NEGF method. A closed-form solution has been obtained and formulated in terms of steady-state nonequilibrium Green's functions. This solution is applicable in the entire nonlinear quantum transport regime and can be used for further $ab~initio$ studies.\cite{footnote2} As a model application of the general solution, we perform a model calculation on an FM/QD/FM system, where the connection between the single-level QD and the leads is described by a Lorentzian linewidth function. Interestingly, it shows that the transient spin current may vary spatially, causing spin accumulation or depletion in the central region. Remarkably, the transient spin current enhances a lot compared to the dc limit of thermoelectric spin current: the in-plane components ($x,z$) of the spin current increase by 2$\sim$3 orders of magnitude under a temperature difference of 30~K; while, the out-of-plane component of the spin current increases by a few percent. Further analysis shows that the transient spin current near $t=0$ has negligible dependence on temperature, but strongly relies on the QD energy level and the noncollinear angle. The key factor in the transient enhancement of the spin current is the transient nature instead of the temperature gradient. Our studies demonstrate that spin currents can be effectively amplified by thermal switches.



\section*{Acknowledgments}
We gratefully acknowledge financial support by the General Research Fund (Grant No. 17311116), the University Grant Council (Contract No. AoE/P-04/08) of the Government of HKSAR (J.W.), NSF-China [Grant Nos. 11774238 (J.W.) and 11704257 (X.C.)], and by the NSERC of Canada (H.G.). We thank Calcul Qu\'ebec and Compute Canada for the computation facilities.

\section*{References}


\begin{thebibliography}{10}

\bibitem{WolfSA_Science_2001}
S.~A. Wolf, D.~D. Awschalom, R.~A. Buhrman, J.~M. Daughton, S.~von Moln¨¢r,
  M.~L. Roukes, A.~Y. Chtchelkanova, and D.~M. Treger.
\newblock Spintronics: A spin-based electronics vision for the future.
\newblock {\em Science}, 294(5546):1488--1495, 2001.

\bibitem{Kuntal_JPD2014}
Roy Kuntal.
\newblock Ultra-low-energy computing paradigm using giant spin hall devices.
\newblock {\em J. Phys. D: Appl. Phys.}, 47(42):422001, 2014.

\bibitem{ZhangShoucheng_Sci2003}
Shuichi Murakami, Naoto Nagaosa, and Shou-Cheng Zhang.
\newblock Dissipationless quantum spin current at room temperature.
\newblock {\em Science}, 301(5638):1348--1351, 2003.

\bibitem{Schmidt_JPD2005}
G.~Schmidt.
\newblock Concepts for spin injection into semiconductors¡ªa review.
\newblock {\em J. Phys. D: Appl. Phys.}, 38(7):R107, 2005.

\bibitem{Moussy_JPD2013}
Moussy Jean-Baptiste.
\newblock From epitaxial growth of ferrite thin films to spin-polarized
  tunnelling.
\newblock {\em J. Phys. D: Appl. Phys.}, 46(14):143001, 2013.

\bibitem{Slonczewski_PRB_1989}
J.~C. Slonczewski.
\newblock Conductance and exchange coupling of two ferromagnets separated by a
  tunneling barrier.
\newblock {\em Phys. Rev. B}, 39(10):6995--7002, 1989.

\bibitem{AlbertFJ_APL2000}
F.~J Albert, J.~A Katine, R.~A Buhrman, and D.~C Ralph.
\newblock Spin-polarized current switching of a co thin film nanomagnet.
\newblock {\em Appl. Phys. Lett.}, 77(23):3809--3811, 2000.

\bibitem{Choi_PRL2007}
Seokhwan Choi, Hyoung~Joon Choi, Jong~Mok Ok, Yeonghoon Lee, Won-Jun Jang,
  Alex~Taekyung Lee, Young Kuk, SungBin Lee, Andreas~J. Heinrich, Sang-Wook
  Cheong, Yunkyu Bang, Steven Johnston, Jun~Sung Kim, and Jhinhwan Lee.
\newblock Switching magnetism and superconductivity with spin-polarized current
  in iron-based superconductor.
\newblock {\em Phys. Rev. Lett.}, 119(22):227001, 2017.

\bibitem{cxb_PRB2017_STT}
Xiaobin Chen, Chenyi Zhou, Zhaohui Zhang, Jingzhe Chen, Xiaohong Zheng, Lei
  Zhang, Can-Ming Hu, and Hong Guo.
\newblock Enhancing the spin transfer torque in magnetic tunnel junctions by ac
  modulation.
\newblock {\em Phys. Rev. B}, 95(11):115417, 2017.


\bibitem{Bauer_NatMater_2012}
Gerrit E.~W. Bauer, Eiji Saitoh, and Bart~J. van Wees.
\newblock Spin caloritronics.
\newblock {\em Nat. Mater.}, 11(5):391--399, 2012.

\bibitem{Slachter_NatPhy_2010}
A.~Slachter, F.L. Bakker, J.P. Adam, and B.J. Van~Wees.
\newblock Thermally driven spin injection from a ferromagnet into a
  non-magnetic metal.
\newblock {\em Nat. Phys.}, 6(11):879--882, 2010.

\bibitem{Cahill_NatPhys2015}
Gyung-Min Choi, Chul-Hyun Moon, Byoung-Chul Min, Kyung-Jin Lee, and David~G.
  Cahill.
\newblock Thermal spin-transfer torque driven by the spin-dependent seebeck
  effect in metallic spin-valves.
\newblock {\em Nat. Phys.}, 11:576, 2015.

\bibitem{Joseph_PRB_2006}
Joseph Maciejko, Jian Wang, and Hong Guo.
\newblock Time-dependent quantum transport far from equilibrium: An exact
  nonlinear response theory.
\newblock {\em Phys. Rev. B}, 74(8):085324, 2006.

\bibitem{WangB_PRB_2010}
Bin Wang, Yanxia Xing, Lei Zhang, and Jian Wang.
\newblock Transient dynamics of molecular devices under a steplike pulse bias.
\newblock {\em Phys. Rev. B}, 81(12):121103, 2010.

\bibitem{zhanglei2012}
Lei Zhang, Yanxia Xing, and Jian Wang.
\newblock First-principles investigation of transient dynamics of molecular
  devices.
\newblock {\em Phys. Rev. B}, 86(15):155438, 2012.

\bibitem{ZhangLei_PRB2013dftTransient}
Lei Zhang, Jian Chen, and Jian Wang.
\newblock First-principles investigation of transient current in molecular
  devices by using complex absorbing potentials.
\newblock {\em Phys. Rev. B}, 87(20):205401, 2013.

\bibitem{ZhouChenyi_PRB2016}
Chenyi Zhou, Xiaobin Chen, and Hong Guo.
\newblock Theory of quantum transport in disordered systems driven by voltage
  pulse.
\newblock {\em Phys. Rev. B}, 94(7):075426, 2016.

\bibitem{Koopmans_NatComm_2014}
AJ~Schellekens, KC~Kuiper, RRJC de~Wit, and B~Koopmans.
\newblock Ultrafast spin-transfer torque driven by femtosecond pulsed-laser
  excitation.
\newblock {\em Nat. Commun.}, 5, 2014.


\bibitem{Zhizhou2017}
Zhizhou Yu, Lei Zhang, and Jian Wang.
\newblock First-principles investigation of transient spin transfer torque in
  magnetic multilayer systems.
\newblock {\em Phys. Rev. B}, 96(7):075412, 2017.

\bibitem{Jauho_PRB_1994}
Antti-Pekka Jauho, Ned~S. Wingreen, and Yigal Meir.
\newblock Time-dependent transport in interacting and noninteracting
  resonant-tunneling systems.
\newblock {\em Phys. Rev. B}, 50(8):5528--5544, 1994.

\bibitem{WangJiansheng_PRB2010}
Eduardo~C Cuansing and Jian-Sheng Wang.
\newblock Transient behavior of heat transport in a thermal switch.
\newblock {\em Phys. Rev. B}, 81(5):052302, 2010.

\bibitem{Gaomin_NJP2017}
Gaomin Tang, Zhizhou Yu, and Jian Wang.
\newblock Full-counting statistics of energy transport of molecular junctions
  in the polaronic regime.
\newblock {\em New J. Phys.}, 19(8):083007, 2017.

\bibitem{cxb_prb2013_ac}
Xiaobin Chen, Dongping Liu, Wenhui Duan, and Hong Guo.
\newblock Photon-assisted thermoelectric properties of noncollinear spin
  valves.
\newblock {\em Phys. Rev. B}, 87(8):085427, 2013.

\bibitem{Zhaohui}
Zhaohui Zhang, Lihui Bai, Xiaobin Chen, Hong Guo, X.~L. Fan, D.~S. Xue,
  D.~Houssameddine, and C.~M. Hu.
\newblock Observation of thermal spin-transfer torque via ferromagnetic
  resonance in magnetic tunnel junctions.
\newblock {\em Phys. Rev. B}, 94(6):064414, 2016.

\bibitem{footnote1}
Since interaction between the leads and the central region exists only after
  $t=0$, self-energies are nonzero only when $t, t_1>0$. Also, because the
  Green's functions for disconnected individual parts ($ g ^\gamma $) are
  time-translational invariant, Fourier transform can be performed to get $g
  ^\gamma \left( {{t_1},t} \right) = \int {\frac{{d\varepsilon }}{{2\pi }}} g
  ^\gamma \left( \varepsilon \right){e^{ - {\rm{i}}\varepsilon \left( {{t_1} -
  t} \right)}}$.

\bibitem{Jauho_Book}
Hartmut Haug and Antti-Pekka Jauho.
\newblock {\em Quantum kinetics in transport and optics of semiconductors}.
\newblock Springer-Verlag, Berlin, 1996.

\bibitem{jiangtao}
Jiangtao Yuan.
\newblock {\em Investigations Of Time-Dependent Quantum Transport Properties In
  Nanoscale Structures}.
\newblock PhD thesis, The University of Hong Kong, Hong Kong, 2017.

\bibitem{TangGaomin_prb2014}
Gaomin Tang and Jian Wang.
\newblock Full-counting statistics of charge and spin transport in the
  transient regime: A nonequilibrium green's function approach.
\newblock {\em Phys. Rev. B}, 90(19):195422, 2014.

\bibitem{Stefanucci_PRB2004}
Gianluca Stefanucci and Carl-Olof Almbladh.
\newblock Time-dependent partition-free approach in resonant tunneling systems.
\newblock {\em Phys. Rev. B}, 69(19):195318, 2004.

\bibitem{Datta_2005_transistor}
Supriyo Datta.
\newblock {\em Quantum transport: atom to transistor}.
\newblock Cambridge University Press, Cambridge, 2005.

\bibitem{Swirkowicz_PRB_2009}
R.~Swirkowicz, M.~Wierzbicki, and J.~Barnas.
\newblock Thermoelectric effects in transport through quantum dots attached to
  ferromagnetic leads with noncollinear magnetic moments.
\newblock {\em Phys. Rev. B}, 80(19):195409, 2009.

\bibitem{Tang_Arxiv2017}
Gaomin Tang, Fuming Xu, Shuo Mi, and Jian Wang.
\newblock {\em arXiv:1712.00215}.

\bibitem{Jauho_PRB2009}
Troels Markussen, Antti-Pekka Jauho, and Mads Brandbyge.
\newblock Electron and phonon transport in silicon nanowires: Atomistic
  approach to thermoelectric properties.
\newblock {\em Phys. Rev. B}, 79(3):035415, 2009.

\bibitem{cxb_PRB2015valley}
Xiaobin Chen, Lei Zhang, and Hong Guo.
\newblock Valley caloritronics and its realization by graphene nanoribbons.
\newblock {\em Phys. Rev. B}, 92(15):155427, 2015.

\bibitem{Datta_prb_2003}
Magnus Paulsson and Supriyo Datta.
\newblock Thermoelectric effect in molecular electronics.
\newblock {\em Phys. Rev. B}, 67(24):241403, 2003.

\bibitem{Dattabook}
Supriyo Datta.
\newblock {\em Electronic transport in mesoscopic systems}.
\newblock Cambridge University Press, UK, 1997.

\bibitem{Bulter_PRB2001}
W.~H. Butler, X.~G. Zhang, T.~C. Schulthess, and J.~M. MacLaren.
\newblock Spin-dependent tunneling conductance of Fe$|$MgO$|$Fe sandwiches.
\newblock {\em Phys. Rev. B}, 63(5):054416, 2001.

\bibitem{footnote2}
Besides the multi-level QD system investigated, our formulas [Eqs.~(\ref{eq:JLa_spin}-\ref{eq:C(E,t)})] are also applicable for multi-site
  systems.

\end{thebibliography}

\appendix

\section{General expression for time-dependent spin currents}
       The general form of a spin current flowing from the central region to lead $l$ is ($\hbar=1$)\cite{cxb_PRB2017_STT}
        \begin{align}\label{app:Jspin}
        {\bf{J}}_l^{spin}\left( t \right)
        &= -  \sum\limits_{ss's''\atop k\alpha \in l,  n  \in C} {\mathop{\rm Re}\nolimits} \left[ {{{\bm{\sigma }}_{s',s''}}{t_{k\alpha s'',n s}}G_{n s,k\alpha s'}^ < \left( {t,t} \right)} \right],
        \end{align}
      where ${\bf{J}}^{spin}=(J_{x}^{spin},J_{y}^{spin}, J_{z}^{spin})$, ${\bm{\sigma }} =(\sigma_x,\sigma_y,\sigma_z)$, $\sigma_{x/y/z}$ are Pauli matrices, and the lesser Green's function is defined as\cite{Jauho_PRB_1994}
       \begin{align}
        G_{ns,kas'}^ < \left( {t,t'} \right) &= \textrm{i}\langle {c_{k\alpha s'}^ \dagger \left( {t'} \right)d_{ns} \left( t \right)} \rangle.
        \end{align}
      Assuming that the hopping between lead $l$ and the central region does not cause spin-flipping, $i.e.$, ${t_{is'',js}}={t_{i,j}\delta_{ss''}}$ is spin-independent, we have
      \begin{align}
      {\bf{J}}_{l}^{spin}(t) =  - \sum\limits_{ss' \atop k\alpha  \in l,n \in {\rm{C}}} {{\rm{Re}}\left[ {{{\bm{\sigma }}_{s's}}{t_{k\alpha ,n}}\left( t \right)G_{ns,k\alpha s'}^ < \left( {t,t} \right)} \right]}.
      \end{align}
      By analytic continuation rules\cite{Jauho_PRB_1994}, the lesser Green's function $ G_{ns,k\alpha s'}^ < \left( {t,t} \right)$ can be written
      in terms of Green's functions of leads in the uncoupled system ($g_{k\alpha s,k' \alpha'  s'}^ \gamma$) and Green's functions of the central region in the coupled systems ($ G_{ns,n's'}^\gamma$). Then, the spin current in lead $l$ turns to be
      \begin{align}\label{app:Jspinl_detailed}
     & {\bf{J}}_{l}^{spin}(t) =   - \sum\limits_{ss's'' \atop nn'n'' \in  C } \int   {d{t_1}} {\rm{Re}}\left\{ {{\bm{\sigma }}_{s's}} \left[ G_{ns,n''s''}^r\left( {t,{t_1}} \right) \cdot     \right. \right. \cr
        &  \Sigma _{l;n''s'',ns'}^ < \left( {{t_1},t} \right) +
       \left.\left. {G_{ns,n''s''}^ < \left( {t,{t_1}} \right)\Sigma _{l;n''s'',ns'}^a\left( {{t_1},t} \right)} \right]\right\},
        \end{align}
       where the time-dependent self-energies of lead $l$ are defined as ($\gamma=>,<,r,a$)
       \begin{align}
       \Sigma _{l;n''s'',ns'}^ \gamma \left( {{t_1},t} \right) = &\sum\limits_{k''\alpha '',k\alpha  \in l}   {t_{n'',k''\alpha ''}}\left( {{t_1}} \right) \cdot \cr
       & g_{k''\alpha ''s'',k\alpha s'}^ \gamma \left( {{t_1},t} \right){t_{k\alpha ,n}}\left( t \right).
       \end{align}
       Written in matrix form, the $\alpha$-component of the spin current flowing through lead $l$ [Eq.~(\ref{app:Jspinl_detailed})] is ($\alpha=x,y,z$)
       \begin{align}\label{app:Jspinl}
        J_{l;\alpha} ^{spin}(t) &=  - \int   {d{t_1}} {\rm{TrRe}}\left\{ {\sigma _\alpha }{{\bf{G}}^r}\left( {t,{t_1}} \right){\bf{\Sigma }}_l^ < \left( {{t_1},t} \right)  \right. \cr
        &+\left.{\sigma _\alpha }{{\bf{G}}^ < }\left( {t,{t_1}} \right){\bf{\Sigma }}_l^a\left( {{t_1},t} \right) \right\}\cr
        &=- \int   {d{t_1}} {\rm{TrRe}}\left\{ {{\bf{G}}^r}\left( {t,{t_1}} \right){\bf{\Sigma }}_l^ < \left( {{t_1},t} \right){\sigma _\alpha } \right. \cr
        &\left.+ {{\bf{G}}^ < }\left( {t,{t_1}} \right){\bf{\Sigma }}_l^a\left( {{t_1},t} \right){\sigma _\alpha } \right\},
        \end{align}
       where the trace goes over both the spin and orbital degrees of freedom. Here and hereinafter, bold-face quantities are defined in the central region.

\section{${\bf{G}}^r(t,t')={\bf{\bar{G}}}^r(t-t')$ ($t,t'>0$) under a sudden thermal switch\label{app:proof}}
   To solve out the retarded Green's function, we introduce a double-time Fourier transform of a function $F(t,t')$ as
        \begin{align}
        F\left( {\varepsilon ,\varepsilon '} \right) = \int_0^{ + \infty } {dt} \int_0^{ + \infty } {dt'F\left( {t,t'} \right){e^{\textrm{i}\varepsilon t  }}{e^{ - \textrm{i}\varepsilon 't'}}}.
        \end{align}
        Note that the integrals of time goes from 0 to $+\infty$, different to the traditional form that goes from $-\infty$ to $+\infty$\cite{cxb_prb2013_ac}. Due to this difference, its inverse transformation is $F(t,t')$ only when both times variables are later than 0:
        \begin{align}
        &\int_{ - \infty }^{ + \infty } {\frac{{d\varepsilon} }{{2\pi }}} \int_{ - \infty }^{ + \infty } {\frac{{d\varepsilon '}}{{2\pi }}} F\left( {\varepsilon ,\varepsilon '} \right){e^{ - \textrm{i}\varepsilon t  }}{e^{\textrm{i}\varepsilon 't'}}\cr
         =& \int   {\frac{{d\varepsilon} }{{2\pi }}} \int   {\frac{{d\varepsilon '}}{{2\pi }}} \int_0^{ + \infty } {du} \int_0^{ + \infty } du' \cdot \cr
        & F\left( {u,u'} \right){e^{\textrm{i}\varepsilon u}}{e^{ - \textrm{i}\varepsilon 'u'}} {e^{ - \textrm{i}\varepsilon t  }}{e^{\textrm{i}\varepsilon 't'}}\cr
         = &\int   {\frac{{d\varepsilon} }{{2\pi }}} \int   {\frac{{d\varepsilon '}}{{2\pi }}} \int_0^{ + \infty } {du} \int_0^{ + \infty } du' \cdot \cr & F\left( {u,u'} \right){e^{\textrm{i}\varepsilon \left( {u - t} \right)}}{e^{ - \textrm{i}\varepsilon '\left( {u' - t'} \right)}} \cr
         =& \int_0^{ + \infty } {du} \int_0^{ + \infty } {du'F\left( {u,u'} \right)\delta \left( {u - t} \right)\delta \left( {u' - t'} \right)} \cr
         =& F\left( {t,t'} \right)\quad \quad t,t' > 0.
        \end{align}
        Under this transform, we may write
        \begin{align}
        {{\bf{G}}^r}\left( {\varepsilon ,\varepsilon '} \right) &= \int_0^{ + \infty } {dt} \int_0^{ + \infty } {dt'{{\bf{G}}^r}\left( {t,t'} \right){e^{\textrm{i}\varepsilon t  }}{e^{ - \textrm{i}\varepsilon 't'}}}, \\
        {{\bf{G}}^r}\left( {t,t'} \right) &= \int_{ - \infty }^{ + \infty } {\frac{{d\varepsilon} }{{2\pi }}} \int_{ - \infty }^{ + \infty } {\frac{{d\varepsilon '}}{{2\pi }}} {{\bf{G}}^r}\left( {\varepsilon ,\varepsilon '} \right){e^{ - \textrm{i}\varepsilon t  }}{e^{\textrm{i}\varepsilon 't'}} \left( {t,t' > 0} \right).
        \end{align}
        \begin{figure}
        \centering
          \includegraphics[width=0.45\textwidth,bb=0 0 440 180]{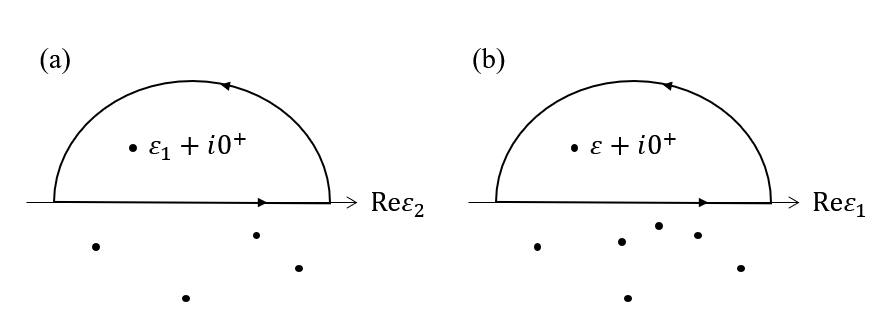}\\
          \caption{Schematic plots of integration contours over $\varepsilon_2 $ and $\varepsilon _3$ in Eq.~(\ref{app:X1}). Black dots in the lower half of the complex plane in (a) and (b) represent poles of ${\bf G}_0^r(\varepsilon )$ and ${\bf G}_0^r(\varepsilon ){\bf \Sigma}^r(\varepsilon )$, respectively.}\label{app:figIntegration}
        \end{figure}
        To work out ${{\bf{G}}^r}$ from the Dyson equation [Eq.~(\ref{app:dyson})], we replace ${{\bf{G}}^r}$ in the r.h.s. of Eq.~(\ref{app:dyson}) by the r.h.s. of Eq.~(\ref{app:dyson}) iteratively, obtaining
        \begin{align}\label{app:Gr=G0r+X1+X2}
        {{\bf{G}}^r}\left( {t,t'} \right) = {\bf{G}}_0^r\left( {t,t'} \right) + {{\bf{X}}_1}\left( {t,t'} \right) + {{\bf{X}}_2}\left( {t,t'} \right) +  \cdots
        \end{align}
        where
        \begin{align}
        {{\bf{X}}_1}\left( {t,t'} \right) =& \int_0^{ + \infty } {d{t_1}} \int_0^{ + \infty } {d{t_2}} \cdot \cr
         & {\bf{G}}_0^r\left( {t,{t_1}} \right){{\bf{\Sigma }}^r}\left( {{t_1},{t_2}} \right){\bf{G}}_0^r\left( {t,t'} \right),\\
        {{\bf{X}}_{n + 1}}\left( {t,t'} \right) = &\int_0^{ + \infty } {d{t_1}} \int_0^{ + \infty } {d{t_2}}\cdot \cr
         &{{\bf{X}}_n}\left( {t,{t_1}} \right){{\bf{\Sigma }}^r}\left( {{t_1},{t_2}} \right){\bf{G}}_0^r\left( {{t_2},\;t'} \right).
        \end{align}
        By utilizing Eqs.~(\ref{app:sigmar_Fourier})(\ref{app:G0r(E)}), ${{\bf{X}}_1}\left( {t,t'} \right)$ can be transformed to
        \begin{align}\label{app:X1}
        &{{\bf{X}}_1}\left( {t,t'} \right)\cr
        =&\int_0^{ + \infty } {d{t_1}} \int_0^{ + \infty } {d{t_2}} \int_{ - \infty }^{ + \infty } {\frac{{d\varepsilon} }{{2\pi }}} {\bf{G}}_0^r\left( \varepsilon \right){e^{-\textrm{i}\varepsilon \left( {t - {t_1}} \right)}} \cdot\cr
        &\int   {\frac{{d{\varepsilon_1 }}}{{2\pi }}} {{\bf{\Sigma }}^r}\left( {{\varepsilon_1 }} \right){e^{-\textrm{i}{\varepsilon_1 }\left( {{t_1} - {t_2}} \right)}}\int   {\frac{{d{\varepsilon_2 }}}{{2\pi }}} {\bf{G}}_0^r\left( {{\varepsilon_2 }} \right){e^{-\textrm{i}{\varepsilon_2 }\left( {{t_2} - t'} \right)}}\cr
        =& \int_{ - \infty }^{ + \infty } {\frac{{d\varepsilon} }{{2\pi }}} {\bf{G}}_0^r\left( \varepsilon \right){e^{-\textrm{i}\varepsilon t  }}\int   {\frac{{d{\varepsilon_1 }}}{{2\pi }}} \frac{\textrm{i}}{{\varepsilon - {\varepsilon_1 } + \textrm{i}{0^ + }}}{{\bf{\Sigma }}^r}\left( {{\varepsilon_1 }} \right) \cdot \cr
        &\int   {\frac{{d{\varepsilon_2 }}}{{2\pi }}} \frac{{\textrm{i}{e^{\textrm{i}{\varepsilon_2 }t'}}}}{{{\varepsilon_1 } - {\varepsilon_2 } + \textrm{i}{0^ + }}}{\bf{G}}_0^r\left( {{\varepsilon_2 }} \right)\cr
        =& \int_{ - \infty }^{ + \infty } {\frac{{d\varepsilon} }{{2\pi }}} {\bf{G}}_0^r\left( \varepsilon \right){e^{-\textrm{i}\varepsilon t  }}\int   {\frac{{d{\varepsilon_1 }}}{{2\pi }}} \frac{{\textrm{i}{e^{\textrm{i}{\varepsilon_1 }t'}}}{{\bf{\Sigma }}^r}\left( {{\varepsilon_1 }} \right){\bf{G}}_0^r\left( {{\varepsilon_1 }} \right)}{{\varepsilon - {\varepsilon_1 } + \textrm{i}{0^ + }}}\cr
        =&\int_{ - \infty }^{ + \infty } {\frac{{d\varepsilon} }{{2\pi }}} {\bf{G}}_0^r\left( \varepsilon \right){{\bf{\Sigma }}^r}\left( \varepsilon \right){\bf{G}}_0^r\left( \varepsilon \right){e^{-\textrm{i}\varepsilon\left( {t - t'} \right)}},
        \end{align}
        where
        \begin{align}
        \int_0^{ + \infty } {{e^{\textrm{i}\omega t}}} dt = \frac{\textrm{i}}{{\omega  + \textrm{i}{0^ + }}}
        \end{align}
        is used, and the integrals over $\varepsilon_1 $ and $\varepsilon_2 $ are carried out in the upper half of complex plane as shown in Fig.~\ref{app:figIntegration} using the theorem of residue. From this result, we know that ${{\bf{X}}_1}\left( {t,t'} \right) $ actually depends on time difference when $t,t'>0$:
        \begin{align}
        {{\bf{X}}_1}\left( {t,t'} \right) = {{\bf{X}}_1}\left( {t - t'} \right)  \quad \quad (t,t'>0)
        \end{align}
        with Fourier components
        \begin{align}
        {{\bf{X}}_1}\left( \varepsilon \right) = {\bf{G}}_0^r\left( \varepsilon \right){{\bf{\Sigma }}^r}\left( \varepsilon \right){\bf{G}}_0^r\left( \varepsilon \right).
        \end{align}
        Similarly,
        \begin{align}\label{app:X2}
        &{{\bf{X}}_2}\left( {t,t'} \right)\cr
        &= \int_0^{ + \infty } {d{t_3}} \int_0^{ + \infty } {d{t_4}} {{\bf{X}}_1}\left( {t - {t_3}} \right){{\bf{\Sigma }}^r}\left( {{t_3},{t_4}} \right){\bf{G}}_0^r\left( {{t_4},\;t'} \right)\cr
        & = \int_{ - \infty }^{ + \infty } {\frac{{d\varepsilon} }{{2\pi }}} {X_1}\left( \varepsilon \right){{\bf{\Sigma }}^r}\left( \varepsilon \right){\bf{G}}_0^r\left( \varepsilon \right){e^{-\textrm{i}\varepsilon\left( {t - t'} \right)}}\cr
        & = {{\bf{X}}_2}\left( {t - t'} \right)
        \end{align}
        with Fourier components
        \begin{align}{{\bf{X}}_2}\left( \varepsilon \right) = {\bf{X}}_1\left( \varepsilon \right){{\bf{\Sigma }}^r}\left( \varepsilon \right){\bf{G}}_0^r\left( \varepsilon \right).
        \end{align}
        Consequently, for arbitrary integer $n>0$ and time $t,t'>0$, we have
        \begin{align}
        {{\bf{X}}_n}\left( {t,t'} \right)& = {{\bf{X}}_n}\left( {t - t'} \right),\\
        {{\bf{X}}_n}\left( \varepsilon \right) &= {{\bf{X}}_{n - 1}}\left( \varepsilon \right){{\bf{\Sigma }}^r}\left( \varepsilon \right){\bf{G}}_0^r\left( \varepsilon \right),
        \end{align}
        where  ${{\bf{X}}_0}\left( \varepsilon  \right)$ is defined as
        \begin{align}
                {{\bf{X}}_0}\left( \varepsilon  \right)& ={\bf{G}}_0^r\left( \varepsilon \right)
        \end{align}
        for convenience.\\

        Therefore, the Dyson equation for the retarded Green's function in Eq.~(\ref{app:dyson}) becomes
        \begin{align}\label{app:Gr=GbarR}
        {{\bf{G}}^r}\left( {t,t'} \right)
        & = {\bf{G}}_0^r\left( {t - t'} \right) + {{\bf{X}}_1}\left( {t - t'} \right) + {{\bf{X}}_2}\left( {t - t'} \right) +  \cdots \cr
        & = \int_{}^{} {\frac{{d\varepsilon} }{{2\pi }}\sum\limits_{n = 0}^{ + \infty } {{\bf{G}}_0^r\left( \varepsilon \right){{\left[ {{{\bf{\Sigma }}^r}\left( \varepsilon \right){\bf{G}}_0^r\left( \varepsilon \right)} \right]}^n}} } {e^{-\textrm{i}\varepsilon\left( {t - t'} \right)}}\cr
        & = \int_{}^{} {\frac{{d\varepsilon} }{{2\pi }}{\bf{\bar G}}^r\left( \varepsilon \right)} {e^{-\textrm{i}\varepsilon\left( {t - t'} \right)}}\cr
        & = {\bf{\bar G}}^r\left( {t-t'} \right),
        \end{align}
        where ${\bf{G}}_0^r\left( \varepsilon  \right)$ is the retarded Green's function for the disconnected central region,
        \begin{align}
        {\bf{G}}_0^r\left( \varepsilon  \right)={\left[ {\varepsilon  + \textrm{i}\eta  - {{\bf{H}}_0}} \right]^{ - 1}},
        \end{align}
        and ${\bf{\bar G}}^r\left( \varepsilon \right)$ is the retarded Green's function for the connected system in steady state\cite{cxb_prb2013_ac}
        \begin{align}
        {\bf{\bar G}}^r\left( \varepsilon \right) &= {\left[ {\varepsilon + \textrm{i}\eta  - {{\bf{H}}_0} - {\bf{\Sigma }}} \right]^{ - 1}}\cr
        & = {\bf{G}}_0^r\left( \varepsilon \right)\sum\limits_{n = 0}^{ + \infty } {{{\left[ {{{\bf{\Sigma }}^r}\left( \varepsilon \right){\bf{G}}_0^r\left( \varepsilon \right)} \right]}^n}}.
        \end{align}


\section{${\bf A}(\varepsilon,t)$}
        For simplifying the expression for spin currents, we introduce the ${\bf A}(\varepsilon,t)$ function as\cite{jiangtao}
        \begin{align}\label{app:defA}
        {\bf{A}}\left( {\varepsilon ,t} \right) = \int_0^t {dt'{{\bf{G}}^r}\left( {t,t'} \right){e^{\textrm{i}\varepsilon \left( {t - t'} \right)}}}.
        \end{align}
        Its Fourier transformation is
        \begin{align}
        \label{eq:Gr=intA}
        {{\bf{G}}^r}\left( {t,t'} \right) = \int_{ - \infty }^{ + \infty } {\frac{{d\varepsilon }}{{2\pi }}{\bf{A}}\left( {\varepsilon ,t} \right){e^{-\textrm{i}\varepsilon \left( {t - t'} \right)}}} \quad t,t' > 0.
        \end{align}
        Using Eq.~(\ref{eq:Gr=GbarR}), ${\bf{A}}\left( {\varepsilon ,t} \right) $ can be rewritten as\cite{Joseph_PRB_2006,jiangtao}
        \begin{align}\label{app:A(E,t)}
        {\bf{A}}\left( {\varepsilon ,t} \right)
        & = {\bf{\bar G}}_{}^r\left( \varepsilon  \right) + \int_{ - \infty }^{ + \infty } {\frac{{d\omega }}{{2\pi i}}\frac{{{e^{-\textrm{i}\left( {\omega  - \varepsilon } \right)t}}}}{{\varepsilon  - \omega  + \textrm{i}{0^ + }}}{\bf{\bar G}}_{}^r\left( \omega  \right)}.
        \end{align}
        This equation offers a simpler way to calculate the ${\bf{A}}\left( {\varepsilon ,t} \right)$.
        Replacing ${\bf \Sigma}^<$ by its Fourier transformation, one finds that the Equation~(\ref{app:keldysh}) can also be rewritten in terms of ${\bf{A}}\left( {\varepsilon ,t} \right)$ as
        \begin{align}\label{eq:G<=ASA}
               {{\bf{G}}^ < }\left( {t,t'} \right) &={{\bf{G}}^r}\left( {t,0} \right){\bf{G}}^ < \left( {0,0} \right){{\bf{G}}^a}\left( {0,t'} \right)+\cr
         \int & {\frac{{d\varepsilon }}{{2\pi }}{\bf{A}}\left( {\varepsilon ,t} \right)} {{\bf{\Sigma }}^ < }\left( \varepsilon  \right){{\bf{A}}^\dagger }\left( {\varepsilon ,t'} \right){e^{-\textrm{i}\varepsilon \left( {t - t'} \right)}}.
        \end{align}
        It is advantageous over the original form in that the double integral is eliminated to a single integral.
\section{Analytical formulas for the spin-degenerate single-level QD with Lorentzian bandwidth functions\label{app:AB}}
The retarded Green's function of the FM/QD/FM system in the steady-state limit when $\theta=0$ is
        \begin{align}
        &{{{\bf{\bar G}}}^r}\left( \varepsilon  \right)\cr
        &={\left[ {\varepsilon  + \textrm{i}\eta   - {\varepsilon _0} - {\bf{\Sigma }}^r} \right]^{ - 1}}\cr
        & = \textrm{diag}\left( {\left[ {\frac{{\varepsilon  + \textrm{i}W}}{{\left( {\varepsilon  - {\omega _{1 \uparrow }}} \right)\left( {\varepsilon  - {\omega _{2 \uparrow }}} \right)}},\frac{{\varepsilon  + \textrm{i}W}}{{\left( {\varepsilon  - {\omega _{1 \downarrow }}} \right)\left( {\varepsilon  - {\omega _{2 \downarrow }}} \right)}}} \right]} \right),\cr
        \end{align}
        where $\omega_{n\sigma}$($n=1,2$; $\sigma=\uparrow\downarrow$) are defined to be poles of the retarded Green's function through factorization of the denominators as:
        \begin{align}\begin{array}{l}
        \left( {\varepsilon  - {\omega _{1 \uparrow }}} \right)\left( {\varepsilon  - {\omega _{2 \uparrow }}} \right) = \left( {\varepsilon  + \textrm{i}W} \right)\left( {\varepsilon  + \textrm{i}\eta   - {\varepsilon _0}} \right) - {\gamma _ \uparrow }W/2,\\
        \left( {\varepsilon  - {\omega _{1 \downarrow }}} \right)\left( {\varepsilon  - {\omega _{2 \downarrow }}} \right) = \left( {\varepsilon  + \textrm{i}W} \right)\left( {\varepsilon  + \textrm{i}\eta   - {\varepsilon _0}} \right) - {\gamma _ \downarrow }W/2
        \end{array}\end{align}
        with $\gamma_\uparrow=\gamma_{L\uparrow}+\gamma_{R\uparrow}$, $\gamma_\downarrow=\gamma_{L\downarrow}+\gamma_{R\downarrow}$.
        Note that due to the retarded nature of ${{{\bf{\bar G}}}^r}$, \{$\omega_{n\sigma}$\} must distribute in the lower half of the complex plane. Bearing this property in mind, we can go forward to get the key functions ${\bf{A}}(\varepsilon,t)$ and ${\bf{B}}_l(\varepsilon,t)$.\\

         By applying the theorem of residue,
       we can obtain analytical expressions for ${\bf{A}}(\varepsilon,t)$:
       \begin{align}
       {\bf{A}}\left( {\varepsilon ,t} \right) &= {\bf{\bar G}}^r\left( \varepsilon  \right) \cr
       &+ \textrm{diag}\left( \left[ {\sum\limits_{n = 1,2}^{} {\frac{{{{{\left( { - 1} \right)}^n}e^{-\textrm{i}\left( {{\omega _{n \uparrow }} - \varepsilon } \right)t}}\left( {{\omega _{n \uparrow }} + \textrm{i}W} \right)}}{{\left( {\varepsilon  - {\omega _{n \uparrow }} } \right)\left( {{\omega _{1 \uparrow }} - {\omega _{2 \uparrow }}} \right)}},}}\right.\right. \cr
       &\left. \left.{{\sum\limits_{n = 1,2}^{} {\frac{{{{{\left( { - 1} \right)}^n}e^{-\textrm{i}\left( {{\omega _{n \downarrow }} - \varepsilon } \right)t}}\left( {{\omega _{n \downarrow }} + \textrm{i}W} \right)}}{{\left( {\varepsilon  - {\omega _{n \downarrow }} } \right)\left( {{\omega _{1 \downarrow }} - {\omega _{2 \downarrow }}} \right)}}} } } \right] \right)
       \end{align}
        and ${\bf{B}}_l(\varepsilon,t)$:
        \begin{align}
        {{\bf{B}}_l}\left( {\varepsilon ,t} \right)
        &= {{{\bf{\bar G}}}^a}{\bf{\Sigma }}_l^a \cr
        &+ \textrm{diag}\left( {\left[ { - \frac{{{\gamma _{l \uparrow }}W}}{2}\sum\limits_{n = 1,2}^{} {\frac{{{e^{\textrm{i}\left( {\omega _{n \uparrow }^* - \varepsilon } \right)t}}}}{{\varepsilon  - \omega _{n \uparrow }^* }}\frac{{{{\left( { - 1} \right)}^n}}}{{\left( {\omega _{2 \uparrow }^* - \omega _{1 \uparrow }^*} \right)}}} ,} \right.} \right.\cr
        &\left. {\left. { - \frac{{{\gamma _{l \downarrow }}W}}{2}\sum\limits_{n = 1,2}^{} {\frac{{{e^{\textrm{i}\left( {\omega _{n \downarrow }^* - \varepsilon } \right)t}}}}{{\varepsilon  - \omega _{n \downarrow }^* }}\frac{{{{\left( { - 1} \right)}^n}}}{{\left( {\omega _{2 \downarrow }^* - \omega _{1 \downarrow }^*} \right)}}} } \right]} \right).
        \end{align}
        In the above, we have acquired explicit analytical expressions for ${\bf{A}}(\varepsilon,t)$ and ${\bf{B}}_l(\varepsilon,t)$. Then, the transient spin current under a sudden thermal switch can be calculated by numerically carrying out one single integral over energy in Eq.~(\ref{eq:JLa_spin}). For cases with $\theta\ne 0$, the above procedures can be performed similarly, except that the number of poles doubles when $\theta\ne 0^\circ$ and $180^\circ$.

\end{document}